\documentclass[11pt]{article}
\usepackage{amsfonts,amssymb,amsmath,graphics,mathtools}
\usepackage{geometry,graphicx, color, psfrag}
\usepackage{tikz, tkz-euclide}
\usetikzlibrary{calc}
\usetikzlibrary{decorations.markings}

\usepackage[english]{babel}
\usepackage{color}

\usepackage{harpoon}
\newcommand*{\nicevec}[1]{\overrightharp{\ensuremath{#1}}}

\usepackage[normalem]{ulem}

\newcommand{\ew}{\color{black}}

\newtheorem{theorem}{Theorem}
\newtheorem{definition}{Definition}
\newtheorem{remark}{Remark}
\newtheorem{proof}{Proof}

\newtheorem{proposition}{Proposition}
\newtheorem{lemma}{Lemma}

\newcommand{\beq}{\begin{eqnarray}}
\newcommand{\eeq}{\end{eqnarray}}

\newcommand{\beqt}{\begin{eqnarray*}}
\newcommand{\eeqt}{\end{eqnarray*}}

\newcommand{\be}{\begin{equation}}
\newcommand{\ee}{\end{equation}}

\newcommand{\bl}{\begin{lemma}}
\newcommand{\el}{\end{lemma}}

\newcommand{\bcon}{\begin{conjecture}}
\newcommand{\econ}{\end{conjecture}}

\newcommand{\br}{\begin{remark}}
\newcommand{\er}{\end{remark}}

\newcommand{\bt}{\begin{theorem}}
\newcommand{\et}{\end{theorem}}

\newcommand{\bd}{\begin{definition}}
\newcommand{\ed}{\end{definition}}

\newcommand{\bp}{\begin{proposition}}
\newcommand{\ep}{\end{proposition}}

\newcommand{\bc}{\begin{corollary}}
\newcommand{\ec}{\end{corollary}}

\newcommand{\bpr}{\begin{proof}}
\newcommand{\epr}{\end{proof}}

\newcommand{\bi}{\begin{itemize}}
\newcommand{\ei}{\end{itemize}}

\newcommand{\ben}{\begin{enumerate}}
\newcommand{\een}{\end{enumerate}}

\usepackage{authblk}
\usepackage{hyperref}
\hypersetup{
    colorlinks=true,
    citecolor=red,
    linkcolor=blue,
    filecolor=magenta,      
    urlcolor=green!80!black,
    pdfpagemode=FullScreen,
    }

\newcommand{\Z}{\mathbb Z}
\newcommand{\R}{\mathbb R}

\newcommand{\s}{\ensuremath{\mathcal{S}}}

\newcommand{\Om}{\ensuremath{\Omega}}

\newcommand{\La}{\ensuremath{\Lambda}}

\DeclareMathOperator{\Hessian}{Hess}

\begin{document}

\title{{\bf  Decimations for One- and Two-dimensional  Ising and Rotator Models II: \\Continuous versus Discrete Symmetries }}
 
\author[1,2]{Matteo D'Achille\thanks{Corresponding author. Email: \href{mailto:email@example.com}{matteo.dachille@universite-paris-saclay.fr}}}
\author[3]{Aernout C.D. van Enter}
\author[2]{Arnaud Le Ny}
\affil[1]{\small Université Paris-Saclay, CNRS, Laboratoire de mathématiques d’Orsay, 91405, Orsay, France}
\affil[2]{\small LAMA UPEC $\&$ CNRS, Universit\'e Paris-Est,  94010 Cr\'eteil, France}
\affil[3]{\small Bernoulli Institute, University of Groningen, 9747AG, Groningen, The Netherlands}

\maketitle
{\bf Abstract:} We show how decimated Gibbs measures which have  an unbroken continuous  symmetry  due to the Mermin-Wagner theorem,  although their discrete equivalents have a phase transition,  still can become non-Gibbsian. The mechanism rests on the occurrence of a spin-flop transition with a broken discrete symmetry, once the model is constrained by the decimated spins in a suitably chosen ``bad'' configuration.

  


\normalsize

\section{Introduction}
In this paper, the second part of our study of decimation for vector models,  following \cite{DDAELN}, we further investigate
the behaviour of spin models under decimations. We show that a number of models which just avoid having spontaneous magnetisation, due to the Mermin-Wagner theorem, still can become non-Gibbsian after decimation at low temperatures. The underlying mechanism is that conditioned on a ``bad'' spin configuration, the conditioned system acquires a discrete symmetry, rather than the original continuous one, and this discrete symmetry  at low temperatures can be broken. This ``spin-flop'' transition is shown to occur in some one-dimensional and two-dimensional examples.

Just as in part I \cite{DDAELN}, our approach will be to prove that a decimated measure will be non-Gibbsian,  when a first-order phase transition {\em in a broad sense} occurs for the ``hidden'' (amongst the initial, non-primed) spins, conditioned on the ``visible'' (decimated, primed) spins to be in a particular (``bad'') configuration. The difference is that here the original model is rotation-invariant and moreover itself has {\em stricto-senso} no such phase transition due to the Mermin-Wagner theorem (abbreviated MW), but by choosing a particular configuration to condition for the visible spins,  this continuous rotation symmetry for the hidden spins gets reduced to a discrete one. For example, in the classical n.n.~$2d$-MW setting of the $XY$ (or plane rotator) model on $\Z^2$, by taking a configuration in which the visible spins point only North or South, the hidden-spin system of the decimation process can become a model in a North-South periodic external field, which can take only those two directions, and the model then is symmetric in a flip of all the components of the spins  in the perpendicular (East-West) direction, which can be spontaneously broken. The difference from our arguments in part I \cite{DDAELN} is that in those cases  the constrained models had similar transitions as the unconstrained models, whereas here the constraints indeed change the symmetry, creating the possibility of a transition  which was excluded in the original models. 

We consider vector spin models of classical two-component unit spins given by an angle $\theta_i \in ]-\pi,  \pi]$ at each site $i  \in \Z^d$, mostly in dimensions $d=1$ or $d=2$.  If $\theta_i =0$, we say that the spin points in the direction East (E); and we say that it points West (W) if $\theta_i = \pi$. Similarly $\theta_i = \frac{\pi}{2}$ denotes North (N) and $ \theta_i = - \frac{\pi}{2} $ South (S). {These are particular instances of $O(N)$ models on $\mathbb{Z}^{2}$, $N=2$, in which the state space is $\Omega_{0}=\left\lbrace v \in \mathbb{R}^N \; \text{s.t.}~||v||_{2}=1 \right\rbrace \equiv \mathbb{S}^{N-1}$, and the finite-volume configuration spaces are $\Omega_{\Lambda}=\Omega_{0}^{\Lambda}$ for any $\Lambda \Subset \mathbb{Z}^{2}$ finite. }
If a ferromagnetic, classical  $XY$ model is subjected to an external field 
which is periodic  or random in the North-South direction, but without having an overall preference {(for example, a centered Gaussian field in the random case)}, there are opposing effects due to the ferromagnetic interaction which pushes all spins to be in the same direction, and the external fields which, if the spins would follow them, would make them point in opposite directions. The combined effect of the two terms is that the ground states and low-temperature Gibbs states can display a broken symmetry, here in the East-West direction.
 
This phenomenon goes under the name of {\em spin-flop mechanism} and has been identified and used before to investigate the non-Gibbsianness of stochastically evolved $XY$ models. For some rigorous versions of these spin-flop results and applications, see~\cite{Craw2011, Craw2014, CR2016, CrawRuszel, EKOR2010,ER2008, ER2009}.

\begin{itemize}
\item{\bf  Long-range 1-dimensional $XY$ model.} Let us first consider a 1-dimensio\-nal rotator model,  
and  assume the interaction decay goes as $1/{r^2}$ ({${r}$ being the distance between two spins}) so there is no magnetisation due to a Mermin-Wagner 
argument \cite{Sim81}, whereas a discrete symmetry can be broken with this decay \cite{ACCN,FrSp82,CFMP}.

Take for the decimated-spin configuration an {infinite} doubly periodic one:
\begin{center}
..................NNSSNNSS...................
\end{center}
 In terms of decimated variables $i'= 2i$, for all $i'$,  when $i$ is of the form  $ i={4k}$ or $ i={4k}+1$, for $k \in \Z$, we have that $\theta'_{i'}= \theta_{2i}=+\frac{\pi}{
2}$ while  for all $i'=2i$ such that $i={4k}+2$ or $i= {4k}+3$, for $k \in \Z$, we have that $ \theta'_{i'} =  \theta_{2i}= - \frac{\pi}{
2}$.

Then the invisible spins live on three types of sites, depending {on} whether the site is in between two North prescriptions (type 1), two South ones (type 2) or between a North and a South one (type 3).
{
\begin{itemize}
\item[-] On sites of type 1 and type 2,  a site is between two neighbours pointing in the same direction, whether North or South, and the spin as a consequence experiences an external field 
of strength $|H|=2 \sum_{j=1... \infty} [J(4j-3) - J(4j-1)]$, in the direction in which its neighbours point\footnote{Notice that  $|H|=2 \sum_{j=1}^\infty \left(\frac{1}{(4j-3)^{2}} - \frac{1}{(4j-1)^{2}}\right)=2 G$, where $G=0.915966\ldots$ is Catalan's constant.}; \item[-] On sites of type 3, which are between a North-pointing spin and a South-pointing spin, there is a perfect cancellation and there the spins experience no external field.
\end{itemize}
So the periodic field is of the form {$...0,H,0,-H,0,H,0,-H,0...$}.
}

\medskip

\item{\bf The two-dimensional case: the doubly alternating configuration.} For the two-dimensional {nearest-neighbour} model, if we choose a doubly alternating configuration of visible, decimated spins  (see Fig.~\ref{fig:dblyalt}, black and white balls), there are more (in fact, seven) types of invisible sites, living on the decorated lattice: a site can be 
\begin{description}
\item {1) }between two N's,
\item {2)} between two S's, 
\item {3)} between an N and an S, 
\item {4)} inside a N-square, 
\item {5)} inside an S-square, 
\item {6)} having two N and two S diagonal neighbours on parallel axes, and 
\item {7)} having  two N-neighbours and two S-neighbours on diagonally opposite sites. 
\end{description}
  \begin{figure}[htbp]
  \centering
  \includegraphics[width=.5\linewidth]{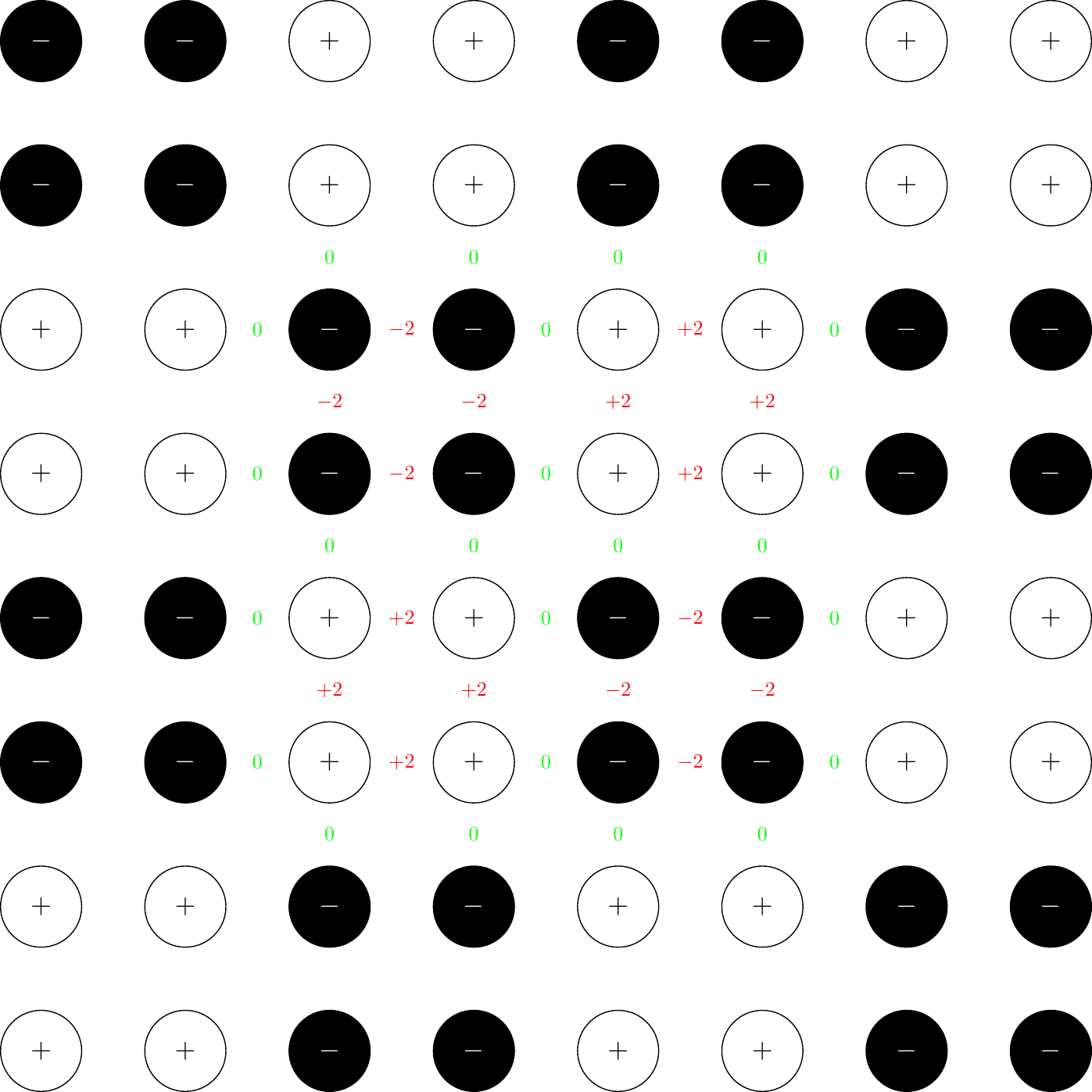}
  \caption{Finite portion of $\sigma_{\rm dblyalt}$.}
  \label{fig:dblyalt}
 \hfill
\end{figure}

 Only the ones between two N's or between two S's experience a field  (see Fig.~\ref{fig:dblyalt}, imbetween sites, in red).
  For $2d$ long-range models also the sites inside an N-square or an S-square {(of so-called types 4 and 5)} feel a field, and a similar analysis applies, with a less diluted periodic external field (non-zero but smaller values for these latter types).

\end{itemize}
We notice that the spin-flop mechanism is fairly robust{:} not only various periodic NS-configurations, but also random ones give rise to it{, so} there are many possible choices for bad configurations {(see e.g.~\cite{Craw2011,Craw2014, CrawRuszel})}. {Notice that} if we would consider decimation not on the even sub-lattice $ (2\Z)^2$, but on a more dilute one, say $(k\Z)^2$,  the ordinary alternating configuration becomes bad, due to this type of  spin-flop argument. {By a similar reasoning}, if we want to have an effectively weak field, one way of achieving this is to consider in each large square a single 2-by-2 square plus or minus, and the rest simply alternating.  

\section{Decimation of the classical XY model}\label{iso-nn-vector}
\subsection{Classical $XY$ model and Mermin-Wagner Theorem}

We consider the case of one of the first  and simplest models of interest  for which a continuous symmetry yields uniqueness in the context of the Mermin-Wagner theorem, the classical (nearest-neighbour) $XY$ model on $\Z^2$, also referred to as the $2d$ n.n.~rotator model.

\subsubsection{Measurable and topological structures} 

Our lattice $S$ is the square lattice $\Z^2$ and the state space is the unit circle $E=\mathbb{S}^{1}$, equipped with the Borel $\sigma$-algebra $\mathcal{E}=\mathcal{B}(\mathbb{S}^{1})$ and with the normalized Haar measure as  {\it a priori} measure $\rho_0 \coloneqq\lambda_{\mathbb{S}^{1}}$, which coincides with the Lebesgue measure on the unit circle. {We denote} by $\left \lbrace\nicevec{e_1}, \nicevec{e_2}\right \rbrace$ the canonical basis of $\mathbb{R}^2$. We also identify $E$ with $]-\pi,+\pi]$ {homeomorphically} : for $i \in \Z^2$, we identify a spin vector $\nicevec{\sigma}_i $ in $E$ by its angle $\theta_i$ w.r.t.\ the horizontal axis,
$$
\theta_i = \theta(\nicevec{\sigma}_i) = (\nicevec{\sigma}_i,\nicevec{e_1}) \in ]-\pi,+\pi]
$$
where $(\cdot , \cdot )$ denotes angle between vectors.

At infinite volume, the {\em configuration space} is the infinite-product probability space 
$$
(\Omega,\mathcal{F},\rho) = (E^{\Z^2},\mathcal{E}^{\otimes \Z^2}, \rho_0^{\otimes \Z^2})
$$
equipped with the product topology. Infinite-volume  configurations are denoted generically by {Greek letters such as} $\nicevec{\sigma}, \nicevec{\omega}$, etc., and form infinite families of {random vectors} ${\bf \nicevec{\sigma}}= (\nicevec{\sigma}_i )_i \in (\mathbb{S}^1)^{\Z^2}$. We also denote $(\Omega_\Lambda, \mathcal{F}_\Lambda, \rho_\Lambda)=(E^\Lambda, \mathcal{E}^{\otimes \Lambda}, \rho^{\otimes \Lambda})$ to be the restriction/projection of $\Omega$ on $\Omega_\Lambda$, for $\Lambda \in \s$, {where the latter is the set of finite subsets of $\Z^2$ (we also sometimes write $\Lambda \Subset \Z^2$ {to stress that the set $\Lambda$ is finite})}. In a similar way, we shall consider possibly infinite subsets $\Delta \subset \Z^2$, {in which case} the preceding notations extend naturally ($\Omega_\Delta,\mathcal{F}_\Delta, \rho_\Delta,\sigma_\Delta$, etc.). 

The configuration space will be also equipped with the product topology for the Borel topology on the sphere $\mathbb{S}^1$ or similarly on the interval $]-\pi,\pi]$. For a given configuration $\nicevec{\omega} \in \Om$, to get  ``open subsets'' with positive measure, {one constrains them to small intervals around  the spin values of a given configuration, rather than requiring a finite number of  spins (here parametrised by their angles)  to be fixed,  as we would do for discrete spins}.  For a given parameter sequence $\epsilon_k >0$, a  basis of open neighbourhoods is thus provided  by configurations that are at most $\epsilon_{k}$-homogeneously ``collinear'' inside {the volume} $\Lambda$, and arbitrary elsewhere. {More formally,} {a basis of neighbourhoods is the family} $\big(\mathcal{N}_{\Lambda,\epsilon_k}(\nicevec{\omega})\big)_{\Lambda \in \mathcal{S}}$ defined by, $\forall \Lambda \in \s$,
$$
\mathcal{N}_{\Lambda,\epsilon_k}(\nicevec{\omega})=\Big \{ \nicevec{\sigma}  \in \Omega :  (\nicevec{\sigma}_i,\nicevec{\omega}_i) {\in (-\epsilon_{k}, +\epsilon_k)}, \forall i \in \Lambda;   \; \nicevec{\sigma}_{\Lambda^c} \; {\rm arbitrary} \Big\}.
$$

{For further {notational} convenience,} we also {introduce} particular open subsets of neighborhoods $\mathcal{N}_{\Lambda,\epsilon} (\nicevec{\omega})$ for which 
the angles are {confined to small intervals of radius $\epsilon >0$ around the maximal and minimal values, so that}  configurations are also close to specific  configurations of canonical angles $0$ or $+\pi$ ($E$ or $W$) on an annulus $\Delta \setminus \Lambda$ for $\Delta \supset \Lambda$. {They are} defined for all $\Lambda \in \s,\; \nicevec{\omega} \in \Om$ as
\begin{eqnarray*}
\mathcal{N}_{\Lambda,\Delta,\epsilon}^{0}(\nicevec{\omega})&\coloneqq & \Big \{ \nicevec{\sigma} \in \mathcal{N}_{\Lambda,\epsilon}(\nicevec{\omega}) : (\nicevec{\sigma}_i, \nicevec{e_1}) \in \Big(-\epsilon,+\epsilon\Big) \; {\rm for} \; i \in {\Delta \setminus \Lambda },\; \nicevec{\sigma}_{\Delta^c} \; {\rm arbitrary} \Big\},\\
\mathcal{N}_{\Lambda,\Delta,\epsilon}^{+\pi}(\nicevec{\omega})&\coloneqq  &\Big \{ \nicevec{\sigma} \in \mathcal{N}_{\Lambda,\epsilon}(\nicevec{\omega}) : (\nicevec{\sigma}_i, \nicevec{e_1})\in \Big( \pi-\epsilon, -\pi +\epsilon \Big) \; {\rm for} \; i \in {\Delta \setminus \Lambda },\; \nicevec{\sigma}_{\Delta^c} \; {\rm arbitrary} \Big\}.
\end{eqnarray*}

We shall sometimes {use the shortcuts} $\mathcal{N}^E \coloneqq \mathcal{N}_{\Lambda,\Delta,\epsilon}^{0}(\nicevec{\omega})$ or $\mathcal{N}^W\coloneqq \mathcal{N}_{\Lambda,\Delta,\epsilon}^{+\pi}(\nicevec{\omega})$.

Similarly, we shall consider neighborhoods $\mathcal{N}^{ME}$ and $\mathcal{N}^{MW}$ in which the spins in the annulus point according to the specific periodic configurations $\theta^{ME}$ -- (by abuse of terminology) called ``Mostly East", alternating NE, E, SE -- or $\theta^{MW}$ -- (similarly) called ``Mostly West", alternating NW, W, SW. These neighborhoods will be used to get an essential discontinuity from the spin-flop transition proved in Section 3.2. 

\subsubsection{Gibbs Measures for vector (rotator) spins}

We consider {\bf Gibbs measures} as macroscopic equilibrium states defined within the DLR framework introduced in the late sixties by Dobrushin~\cite{Dob68}  and Lanford/Ruelle~\cite{LR69} independently. Macroscopic states in general are represented by elements of the set of $\mathcal{M}_1^+$ of probability measures on  $\Omega$. {In the DLR approach} one aims at characterizing the probability measure {\em via} its conditional probabilities w.r.t.\ outsides of finite sets, where boundary conditions are prescribed. Candidates to represent regular systems of conditional probabilities are {called} {\em specifications}, which are families of (proper) probability kernels $\gamma$ introduced by  F\"ollmer~\cite{Foll75} and Preston~\cite{Pres80}  in the late 70's to formalize the following {\em DLR Equation}: A probability measure $\mu$ is said to be specified by $\gamma$, written $\mu \in \mathcal{G}(\gamma)$, when the {elements of the} former represent  versions of these conditional probabilities {w.r.t.\ the outside of finite sets}, {\em i.e.}
\be \label{DLR}
\forall \Lambda \Subset \Z^2, \forall A \in \mathcal{F}, \mu[A \mid \mathcal{F}_{\Lambda^c}](\cdot) = \gamma_\Lambda (A \mid \cdot) \; \mu-{\rm a.s.}\;.
\ee

We emphasize that for a DLR measure $\mu \in \mathcal{G}(\gamma)$, the characterizing DLR equation (\ref{DLR}) is valid for {\em finite} $\Lambda$ only, so in particular this does not provide regular versions of conditional probabilities w.r.t.\ the outside of infinite sets, which are in fact infinite-volume probability measures on a projected configuration space related to an infinite-volume sub-lattice. In this {\em Gibbs vs.\ non-Gibbs framework}, one then needs usually to extend the {(local) specifications $(\gamma_\Lambda)_{\Lambda \in \s}$ into kernels representing versions of conditional probabilities w.r.t.~the outside of infinite sets}, {and this} was precisely the purpose of our {earlier} paper~\cite{DDAELN} in which, among other things, we have described the {\em global specification for ferromagnetic $XY$ models}. {We will make use of it in the present work}.

Gibbs measures are the DLR measures defined for the so-called Gibbs specification $\gamma^{\beta \Phi}$  at inverse temperature $\beta >0$, for an uniformly absolutely convergent potential $\Phi$: For any $\Lambda \Subset \Z^2$, $A \in \mathcal{F}$ and {\em all} boundary condition $\nicevec{\omega} \in \Omega$,
$$
\gamma_\Lambda(A \mid \nicevec{\omega}) = \frac{1}{Z_\Lambda^{\beta \Phi}(\nicevec{\omega})} \int_\Omega \mathbf{1}_A (\nicevec{\sigma}_\Lambda \nicevec{\omega}_{\Lambda^c}) e^{-\beta H_\Lambda(\nicevec{\sigma}_\Lambda \mid \nicevec{\omega}_{\Lambda^c})} \rho_\Lambda \otimes \delta_{\omega_{\Lambda^c}} (d\nicevec{\sigma}),
$$

where $
H_\Lambda (\nicevec{\sigma}_\Lambda | \nicevec{\omega}_{\Lambda^c}) = \sum_{A \cap \Lambda \neq \emptyset} \Phi_A( \nicevec{\sigma}_\Lambda \nicevec{\omega}_{\Lambda^c})
$ is the Hamiltonian\footnote{Which is well defined and finite when the potential is {Uniformly Absolutely Convergent (UAC)}, see \cite{Geo88}.} at finite volume $\Lambda$ with boundary condition $\nicevec{\omega}$  and $Z_\Lambda^{\beta \Phi}(\nicevec{\omega})$ {is} a standard normalization.

We consider {\em Ferromagnetic Pair Potentials}
\be \label{XY}
\Phi_A(\sigma)=\left\{
\begin{array}{lll}
\; - J(i,j) \;\cdot  \langle \nicevec{\sigma}_i ,  \nicevec{\sigma}_j \rangle \; \; & &\textrm{if} \; A=\{i,j\} \;\\
\; 0 \; & &\textrm{otherwise}
\end{array} \right.
\ee
{in which the} coupling function is $J : \Z^2 \times \Z^2 \longrightarrow \R$ and $\langle \ \cdot , \cdot  \ \rangle $ is {\em some} inner product in $\R^2$. We denote more shortly {by} $\gamma^J$ the Gibbsian specification for this potential.  In this section devoted to the classical $XY$ model, the potential is restricted nearest neighbours, and ferromagnetic, that is 
$$
J(i,j)=J \mathbf{1}_{||i-j||_2 =1}, \; J \geq 0.
$$

Before quoting the famous Mermin-Wagner result  proving absence of continuous symmetry breaking in this framework (and which in some cases,  like the standard n.n.\ ferromagnetic $XY$ model, can be strengthened to obtain a form of  uniqueness of Gibbs measures), we investigate the weak limits with prescribed homogeneous b.c., as we shall need them further on. As we shall also see in the next subsection, these models are said to be ferromagnetic because spins have a tendency to align, yielding proper concepts of monotonicity for measures. To express it, we have introduced in \cite{DDAELN} a natural appropriate (but arbitrary) order $\leq_{\rm sin}$, in which configurations compare like $\theta \leq \theta'$ when $\sin{\theta} \leq \sin{\theta'}$. Monotone preservation\footnote{ For this $ $ (pre)-order, or by using alternative correlation inequalities valid for planar $XY$ models, from e.g. \cite{MMSP78, Monroe75}, see also Footnote in Section \ref{Global} or Remark \ref{rkGlo}.} allows indeed to get the well-defined weak limits:

\begin{proposition}\label{Wlimit2}\cite{FrPf81,FrPf83}
Consider the $XY$ models defined {by} (\ref{XY}) with a ferromagnetic coupling function $J(\cdot,\cdot)$, and the boundary condition $\theta=\theta^+={\bf + \frac{\pi}{2}}$ (North) and $\theta=\theta^-={\bf - \frac{\pi}{2}} $ (South). Then the weak limits
\be \label{muplusminus2}
\mu^-(\cdot) \coloneqq   \lim_{\La \uparrow \mathbb{Z}^2} \gamma_\La^J (\cdot | \theta^-)\; \; {\rm and} \; \; \mu^+(\cdot) \coloneqq  \lim_{\La\uparrow \mathbb{Z}^2}  \gamma_\La^J (\cdot | \theta^+)
\ee
are well-defined, translation-invariant and extremal elements of $\mathcal{G}(\gamma^J)$. For any $f$ bounded increasing (for the appropriate $\leq_{\rm sin}$ order), any other measure $\mu \in \mathcal{G}(\gamma^J)$ satisfies 
\be \label{stochdom2}
\mu^-[f] \leq \mu[f] \leq \mu^+[f].
\ee
Moreover, $\mu^-$ and $\mu^+$  are respectively left-continuous and right-continuous.
\end{proposition}

Furthermore, see {\em e.g.} \cite{FrPf81}, it is possible to proceed similarly for \emph{any} homogeneous b.c. ${\bf \theta}$, defined by ${\bf \theta}_i = \theta \in ]-\pi,+\pi]$ at any site $i \in \Z^2$ to define weak limits 
$$
\mu^\theta (\cdot) = \lim_{\Lambda \uparrow Z^2} \gamma_\Lambda^J(\cdot | {\bf \theta})
$$

to eventually get an extremal decomposition in terms of these weak limits: For any ${\bf \theta}$, $\mu^\theta$ is extremal and there is no other extremal state, so that any  $\mu \in \mathcal{G}(\gamma)$ can be formally written
$$
\mu = \int_{-\pi}^{\pi} \alpha_\theta(\mu) d \nu_\theta.
$$

Here $\nu_\theta$ is a measure on $]-\pi,+\pi]$ inheriting from $\mu^\theta$ from the pi-point construction of Georgii, as well as the weights $\alpha_\theta(\mu)$, see~\cite{Geo88}, Chapter 7.

 Nevertheless, in the classical $XY$ model, for n.n.~potentials, all these extremal measures always coincide:
 
\begin{theorem}\label{MW}  I)
{(Mermin-Wagner ban~\cite{MW1966, Mermin67,FV16})}. 

Consider the classical (n.n.) $XY$ model, in dimension two, with n.n.~rotation symmetric pair potentials. Then there is no continuous symmetry breaking. 

II) {(Ferromagnetic uniqueness~\cite {BFL77})}.
 For the standard n.n. ferromagnetic classical  $XY$ model there is a unique translation invariant, extremal  Gibbs measure. For all homogeneous b.c.~$\theta$, the weak limits $\mu^\theta$ coincide at any temperature and thus
$$
\forall \beta >0,\; \mathcal{G}(\gamma^{J})=\{\mu\}
$$
where the unique Gibbs measure $\mu$ can be selected by any of these b.c. (or even with free b.c.).
\end{theorem}

{\bf Quasilocal functions and essential discontinuity:}

In our vector spin context, in contrast  to discrete-spin settings, continuity is not equivalent to quasilocality, but the Gibbsian characterization in terms of non-nullness and quasilocality still holds. Let us recall a few definitions adapted to this context.

A measurable function $f : \Omega \longrightarrow \R^k$ is said to be {\em local} if it is $\mathcal{F}_\Lambda${-measurable} for {some} $\Lambda \Subset \Z^2$. It is said \emph{quasilocal}, and written $f \in \mathcal{F}_{\rm qloc}$, when it is a uniform limit of local functions. Alternatively, {the following characterization holds:}
\be
f \in \mathcal{F}_{\rm qloc} \Longleftrightarrow  \lim_{\Lambda \uparrow \s} \sup_{\nicevec{\sigma}_\Lambda=\nicevec{\omega}_\Lambda} \mid f(\nicevec{\sigma}) - f(\nicevec{\omega}) \mid = 0.
\ee

Quasilocality extends naturally to {specifications and measures: A specification $\gamma$ is quasilocal if its action on local functions $f$ yields quasilocal functions $\gamma_\Lambda f$ for any finite volume $\Lambda$, while a  measure $\mu \in \mathcal{M}_1^+$ is quasilocal if it is consistent with a quasilocal specification $\gamma$}. {Actually}, this quasilocality \emph{completely characterizes} Gibbs measures up to a non-nullness (``finite-energy'') condition, as coined in the mid 70's by Kozlov~\cite{Ko} or Sullivan~\cite{Su} (see also~\cite{BGMMT21}).


\begin{proposition}\cite{Ko,Su,EFS93,LN08}
A   measure   $\mu \in \mathcal{M}_1^+$ is a Gibbs measure iff $\mu$ is quasilocal.
\end{proposition}

In particular, conditional probabilities w.r.t.~the outside of finite sets enjoy nice regularity properties w.r.t.~the boundary condition:  There cannot exist essentially non-local versions of these conditional probabilities {(equivalently, any essentially non-local version of these conditional probabilities is discontinuous)}. Here, by {\em essential} we mean a property that cannot be removed by changes on zero measure sets.

\begin{definition}[Essential discontinuity]
\label{essdiscdef}
A configuration $\nicevec{\omega} \in \Omega$ is said to be a point of essential  discontinuity for a conditional probability of $\mu \in \mathcal{M}_1^+$ if there exists $\Lambda_0 \Subset \Z^2$, $f$ local, $\delta >0$, such that for all $\Lambda$ with $\Lambda_0 \subset \Lambda$ there exist    $\mathcal{N}_\Lambda^1(\nicevec{\omega})$ and $\mathcal{N}_\Lambda^2(\nicevec{\omega})$, two open (or at least positive-measure) neighbourhoods of $\nicevec{\omega}$, such that
$$
\forall \nicevec{\omega}^1 \in \mathcal{N}_\Lambda^1(\nicevec{\omega}),\; \forall \nicevec{\omega}^2 \in \mathcal{N}_\Lambda^2(\nicevec{\omega}), \;\Big| \mu \big[f |\mathcal{F}_{\Lambda^c} \big](\nicevec{\omega}^1) - \mu \big[f |\mathcal{F}_{\Lambda^c} \big](\nicevec{\omega}^2)\Big| > \delta
$$
or equivalently
\be\label{essdisc}
\lim_{\Delta \uparrow \mathbb{Z}} \sup_{\nicevec{\omega}^1,\nicevec{\omega}^2 \in \Omega}  \Big| \mu \big[f |\mathcal{F}_{\Lambda^c} \big](\nicevec{\omega}_\Delta \nicevec{\omega}^1_{\Delta^c}) - \mu \big[f |\mathcal{F}_{\Lambda^c} \big](\nicevec{\omega}_\Delta\nicevec{\omega}^2_{\Delta^c})\Big|  > \delta.
\ee
\end{definition}

It means that regular version of conditional probabilities w.r.t.~the outside of finite sets cannot be changed into a discontinuous version, whatever the boundary condition is.  This {crucial} property will be used throughout this paper to prove non-Gibbsianness of decimated measures: {It suffices to exhibit a point of essential discontinuity, a so-called {\em bad configuration}}.

\subsubsection{Global Specifications}\label{Global}

Let us first prepare the derivation of the conditional probabilities for the decimated measure, before introducing it more formally in next section. As we shall  see, due to the scaling inherent to such  renormalization transformations,  the conditioning w.r.t.~the outside of finite sets for the decimated measure corresponds in terms of the original Gibbs measure to considering  b.c.~outside of a set of infinite size consisting of the locations of the  internal spins (see also \cite{EFS93}). {A {\em local} specification, as described in e.g.~Georgii's book \cite{Geo81}, is not sufficient for this purpose, and in order to proceed in this way} we need to have a global specification. This is exactly what we derived in our first paper on decimation for continuous spins \cite{DDAELN}, where  thanks to monotonicity w.r.t. to this  partial order\footnote{It is in fact, as  a referee pointed out, only a pre-order as ${\rm sin} \; \theta \leq {\rm sin}  \; \theta'$ and ${\rm sin} \  \theta \leq {\rm sin}  \;  \theta'$ does not imply $\theta=\theta'$ in full generality, but this does not affect our results.  The results of Proposition \ref{Wlimit2} are derived in terms of various correlation inequalities valid for planar $XY$ models and similarly  in \cite{DDAELN},    the existence of Global specification  relies on monotonicity arguments that only require the preservation of the pre-order property. 
}
  cooked up there, we indeed proved in~\cite{DDAELN} that for our $XY$ models, it is possible to extend the validity of DLR equations to regular versions of conditional probability w.r.t. arbitrary sets, not necessarily finite.  


{\begin{theorem}\label{globspe-rotator}{(Global specification for $2d$-rotator spins~\cite{DDAELN})}.
Consider any ferromagnetic rotator models on $\Z^2$ defined by (\ref{XY}) at inverse temperature $\beta >0$ with specification $\gamma^J$  with  (ferromagnetic) couplings $J(i,j)$ defined for any pair $\{i,j\} \in \Z^2$, and in particular its  extremal Gibbs measures $\mu^+$ and $\mu^-$,  respectively obtained by weak limits from the opposed angle-b.c.  ${\bf \theta^+} \equiv {\bf +\frac{\pi}{2}}$ or ${\bf \theta^-} \equiv{\bf -\frac{\pi}{2}}$. {Let} $\Gamma^+=(\Gamma_S^+)_{S \subset \Z^2}$ be a family of probability kernels on $(\Omega, \mathcal{F})$  s.t.~:
\begin{itemize}
\item For $S=\Lambda$ finite, for all $\nicevec{\omega} \in \Omega$,
$
\Gamma^+_\Lambda(d \nicevec{\sigma} | \nicevec{\omega}) \coloneqq  \gamma^J_\Lambda (d \nicevec{\sigma} |  \nicevec{\omega}).
$
\item For $S$ infinite, for all $\nicevec{\omega} \in \Omega$,
\be \label{GammaRota}
\Gamma^+_S(d\nicevec{\sigma} | \nicevec{\omega})\coloneqq \mu_S^{+,\nicevec{\omega}} \otimes \delta_{\nicevec{\omega}_{S^c}}(d\nicevec{ \sigma})\; ,
\ee
where the constrained measure $\mu_S^{+,\nicevec{\omega}}$  is the weak limit obtained with freezing in $\nicevec{{\bf +}}_S \nicevec{\omega}_{S^c}$ outside finite volumes: 

$$\mu_S^{+,\nicevec{\omega}}(d \nicevec{\sigma}_S)\coloneqq \lim_{\Delta \uparrow S} \gamma^J_\Delta (d \nicevec{\sigma}\mid \nicevec{{\bf +}}_S \nicevec{\omega}_{S^c}).$$
\end{itemize}
Then $\Gamma^+$ is a global specification such that $\mu^+ \in \mathcal{G}(\Gamma^+)$.  Similarly, one defines a monotonicity-preserving and left-continuous global specification $\Gamma^-$ such that $\mu^- \in \mathcal{G}(\Gamma^-)$.
\end{theorem}}

In our uniqueness framework, these global specifications $\Gamma^+$ and $\Gamma^-$ of course coincide, but we use this more general result in next subsection to prove non-Gibbsianness of the decimated measures, for which conditional probabilities w.r.t.~the outside of finite sets transfer naturally into conditional probabilities w.r.t.~the outside of {\em internal} non-finite sets for the original Gibbs measure. We shall be able to describe an essential ``spin-flop'' discontinuity, leading to non-Gibbsianness of the decimated measure thanks to the characterization of Gibbs measures in terms of quasilocality.

\begin{remark}\label{rkGlo}

The global specification property is a short-cut in these type of arguments which applies to various transformations,  in monotonic situations, as originally wroked out in \cite{FP97} for Ising spins only.  For  decimation transformations we also might have followed the direct analysis  from \cite{EFS93}, which was worked out for decimation transformations, to derive existence of "bad points".

\end{remark}

\subsection{Non-Gibbsianness {\em via} spin-flop}

{\bf {\em Decimated Measures :}}

We start from the unique Gibbs measure $\mu$ for the $XY$ model on $\Z^2$. In particular, $\mu=\mu^+=\mu^-=\mu^\theta$ and can be selected by any of these boundary conditions {(and thus even with free b.c.)}. We shall sometimes write $\mu^+$, as we shall need the global specification $\Gamma^+$ built for it\footnote{The global specification could also be defined in the uniqueness case but we use the construction of \cite{DDAELN} valid in the monotone case, as it already exists for these $2d$-vector spins. Note that we could also modifiy our {\em ad-hoc} order to build a global specification $\Gamma^\theta$ for any of the homogeneous extremal $\mu^\theta$ in case they differ.}. 

 {As in~\cite{DDAELN}}, we shall submit these Gibbs measures to the {\em decimation transformation}  :
\be \label{DefDec}
 T \colon (\Omega,\mathcal{F})  \longrightarrow (\Omega',\mathcal{F}')=(\Omega,\mathcal{F}); \; 
\nicevec{\omega} \; \;   \longmapsto \nicevec{\omega}'=(\nicevec{\omega}'_i)_{i \in
\mathbb{Z}^2}, \; {\rm with} \;  \nicevec{\omega}'_{i}=\nicevec{\omega}_{2i} \; .
\ee
Denote by $\nu  \coloneqq T \mu $ the decimated measure, formally defined as an image measure {\em via}
$$
\forall A' \in \mathcal{F'},\; \nu (A')=\mu (T^{-1} A')=\mu (A) \;, \qquad {\rm where} \; A=T^{-1} A'= \big\{\nicevec{\omega}: \nicevec{\omega}'=T (\nicevec{\omega}) \in A' \big\}.
$$
We distinguish between original {(invisible)} and image {(visible)} sets using primed notation, although by  rescaling {in the case of decimation} the {infinite} configuration spaces $\Omega$ (original) and $\Omega'$ (image) are {actually in bijection}. 
 
  To proceed, we consider some local functions  $f$  and  investigate a potential essential discontinuity by the evaluation of the conditional probabilities
\be \label{condmagn-rota}
 \nu [f(\nicevec{\sigma}')| \mathcal{F}_{\{(0,0)\}^c} ](\nicevec{\omega}') = \mu [f(\nicevec{\sigma}')| \mathcal{F}_{S^c} ](\nicevec{\omega}),\; \nu-{{\rm a.s.}}\; ,
\ee

where $S^c=(2 \mathbb{Z}^2) \cap \{(0,0)\}^c={\left( (2 \mathbb{Z}^2)^c \cup \{(0,0)\}\right)^{c}}$ {\underline{is not finite}}: {\em the conditioning  is {not} on the complement of a finite set}; so that DLR equations do not hold. To circumvent this problem, we use thus the Global Specifications $\Gamma^+$ from~\cite{DDAELN}. As described there and in Theorem \ref{globspe-rotator}, it is defined after a weak limit performed on a constrained local specification  

$$
\gamma^{S,\nicevec{\omega}}=\big(\gamma_{\Lambda}^{S,\nicevec{\omega}}\big)_{\Lambda \in \mathcal{S}}
$$ (for $S=(2\Z)^2$) and a candidate measure in $\mathcal{G}(\gamma^{S,\nicevec{\omega}})$, called the constrained measure $\mu^{+,\nicevec{\omega}}_{S}$, is defined {\em via} the weak limit
\be
\label{eq.wl1235}
\mu^{+,\nicevec{\omega}}_{S} (\cdot) \coloneqq \lim_{\Delta \uparrow S}\gamma_{\Delta}^{J}(\cdot \mid \nicevec{+}_{S}\nicevec{\omega}_{S^{c}}),
\ee
and gives rise, for any infinite set $S\subset \mathbb{Z}^{2}$, to the kernels
$$
\Gamma_{S}^{+}(d\nicevec{\sigma}\mid \nicevec{\omega})\coloneqq\mu^{+,\nicevec{\omega}}_{S} (d\nicevec{\sigma}_{S})  \otimes \delta_{\nicevec{\omega}_{S^c}}(d\nicevec{ \sigma}_{S^{c}})
$$
which may be also written as
$$
\Gamma_{S}^{+}(d\nicevec{\sigma}\mid \nicevec{\omega}) = \lim_{\Delta \uparrow S}\gamma_{\Delta}^{J}(d\nicevec{\sigma} \mid \nicevec{+}_{S}\nicevec{\omega}_{S^{c}}).
$$

Now, for any special  configuration $\nicevec{\omega}'_{\rm spe}$ {(which we shall exhibit later on)}, (\ref{condmagn-rota}) reduces for $\nu$-a.e. $\nicevec{\omega}' \in \mathcal{N}_{\Lambda',\epsilon}(\nicevec{\omega}'_{\rm spe}) $ to
\be \label{condmagn22}
 \nu[f(\nicevec{\sigma}')| \mathcal{F}_{\{(0,0)\}^c} ](\nicevec{\omega}') =  \Gamma_{S}^+ [f(\nicevec{\sigma}')| \nicevec{\omega}] \; \; \mu {\rm -a.e.} (\nicevec{\omega}) \; ,
\ee
with $S =(2 \mathbb{Z}^2)^c \cup \{(0,0)\}$ and {the hidden/invisible configuration} $\nicevec{\omega} \in T^{-1} \{\nicevec{\omega}'\}$ is {some} chosen special configuration on the even lattice $2\Z^2$. The expression of the rhs of (\ref{condmagn22}) is provided in terms of the constrained measure $\mu^{+,\nicevec{\omega}}_{(2\mathbb{Z}^2)^c \cup \{ (0,0)\}}$, with $\nicevec{\omega} \in T^{-1} \{\nicevec{\omega}'\}$  so that we get, for any $\nicevec{\omega}' \in \mathcal{N}_{\Lambda'}(\nicevec{\omega}'_{\rm spe})$,

$$
 \nu[f(\nicevec{\sigma}') | \mathcal{F}_{\{(0,0)\}^c} ](\nicevec{\omega}') =  \mu^{+,\nicevec{\omega}}_{(2\mathbb{Z}^2)^c \cup \{(0,0)\}} \otimes \delta_{\nicevec{\omega}_{2\mathbb{Z}^2 \cap \{(0,0)\}^c}} [f(\nicevec{\sigma}')].
$$
It can be explicitly built as the monotone weak limit   obtained by ${\bf +\frac{\pi}{2}}$-b.c. fixed after a freezing of $\nicevec{\omega}$ on the even sites :  $\forall \nicevec{\omega}' \in \mathcal{N}_{\Lambda'}(\nicevec{\omega}'_{\rm alt}), \forall \nicevec{\omega}\in T^{-1} \{\nicevec{\omega}'\},\;  $
\be \label{constrLimit2}
 \mu^{+,\nicevec{\omega}}_{(2\mathbb{Z}^2)^c \cup \{(0,0)\}} (\cdot) =\lim_{\Delta \in\s,\Delta \uparrow (2 \mathbb{Z}^2)^c  \cup \{0,0)\}} \gamma^J_\Delta (\cdot\mid \nicevec{+}_{(2 \mathbb{Z}^2)^c  \cup \{0,0)\})} \nicevec{\omega}_{2 \mathbb{Z}^2 \cap\{0,0)\}^c}).
\ee


Consider the settings and notations of our first paper (\cite{DDAELN}, e.g.\ Section 5) but this time with an anisotropy parameter $\kappa=0$ we recover the classical $XY$ model where MW holds so that the weak limits $\mu^+=\mu^{+\frac{\pi}{2}}$ and $\mu^-=\mu^{-\frac{\pi}{2}}$ coincide, and thus one cannot get an essential discontinuity as in (5.37) of \cite{DDAELN} for the simply alternating configuration.

Unlike in~\cite{DDAELN}, where the {special/bad} configuration was the alternating configuration $\nicevec{\sigma}'_{\rm alt}$, here we {need to} consider the \emph{doubly alternating configuration} $\nicevec{\sigma}'= \nicevec{\sigma}'_{\rm dblyalt}$. \\
{Recalling that on the decimated lattice a site/vertex has integer coordinates ${i',j'}$, a configuration $\nicevec{\sigma}'$ is doubly alternating if
$$
 \nicevec{\sigma}'_{\{i',j'\}} = (-1)^{\lfloor \frac{i'}{2}\rfloor + \lfloor \frac{j'}{2} \rfloor } \, ,
 $$
where $\lfloor x \rfloor$ is the integer part (floor) of $x$, the largest integer smaller than or equal to $x$.

}




\begin{theorem}[{Essential discontinuity {\em via} spin-flop}]\label{DecXY}

The config\-ura\-tion $\nicevec{\sigma}'_{\rm dblyalt}$ is a bad configuration for the decimated measures $\nu = T\mu$ of the classical $XY$ model {at low enough temperatures}.

\end{theorem}
{\bf Proof.} 
{Let $\nicevec{\omega}' \in \mathcal{N}_{\Lambda, \epsilon}(\nicevec{\sigma}'_{\rm dblyalt})$}.
Then, after decimation on the sublattice $S=(2\Z)^2$, the expectation of any local function $f$ has to be understood as the global specification of a rotator system of the form (\ref{XY}), expressed in terms of the constrained measure (\ref{eq.wl1235}) obtained  by the weak limit (\ref{constrLimit2}).

By imposing the specific constraint on the even lattice, we are driven to investigate a rotator model on the decorated lattice $S$, of formal Hamiltonian on $\Omega_S$
$$
H^{J,NS} \coloneqq \sum_{A \Subset S} \Phi^{J,NS}_A
$$ 
with a pair potential $\Phi^{J,NS}=\big(\Phi^{J,NS}_A\big)_{A \Subset S}$ composed of the initial coupling $J$ for n.n.~pair interaction
\be \label{HamPerio}
\Phi_{\{i,j\}}^{J,NS}(\nicevec{\sigma})  =  -  J \mathbf{1}_{|i-j|=1} \;\cdot  \langle \nicevec{\sigma}_i ,  \nicevec{\sigma}_j \rangle 
\end{equation}
 {\bf plus} an external field $h=(h_i)_{i \in S}$ {acting only on the sites of the decorated lattice $S$ as a self-interaction term, namely} 
 \begin{equation}
 \label{eq.selfint}
 \Phi_{\{i \}}^{J,NS}(\nicevec{\sigma}) = -  h_i   \cdot   \langle \nicevec{\sigma}_i , \nicevec{e_2} \rangle =- h_{i} \cos{\left(\frac{\pi}{2}-\theta_{i}\right)}, \quad i \in S
 \end{equation}
 (and $\Phi^{J,NS}_A = 0$ otherwise). The value of the external field $\left(h_i\right)_{i\in S}$ alternates as follows :

\begin{itemize}
\item On {\em even} vertical levels, for  $i_2=2k,\, k\in \Z $ (where one site out of two has been suppressed to form the decorated lattice), the field alternates between $h_i=+2$ (when located in between two North original spins{, type 1}), $h_i=0$ when the site is between two opposing $N$ and $S$ original spins (type 3 and 4, between $NS$ and $SN$, respectively) and $h_i=-2$ (when located in between two South original spins, {type 2}).

\item On {\em odd} vertical levels, for $i_2=2k+1, \, k\in \Z$ , the external field is diluted to zero: $h_i=0$. This holds for different reasons depending on the horizontal component: On even horizontal levels ({type 2}), this is due to exact cancellation from the opposite $N$ and $S$ vertical neighbours {(type 3, 6 and 7)}, while on odd horizontal levels {(type 4 and 5)}, {where a site is not n.n.~of any of the} $2 \Z^2$, no effect at all of the prescription of the decimated spin is felt. Note that the latter will be a real dilution, but non-null, when considering original systems with ranges longer than n.n.

\end{itemize}

Therefore for a given configuration $\nicevec{\sigma}$ of angles $\theta=(\theta_i)_{i \in \Z^2}$, the value of this self-interaction part, see (\ref{eq.selfint}), on the  sites of the even vertical levels can be written
\begin{equation}\label{Period}
\Phi_{\{i\}}^{J,NS}(\nicevec{\sigma})= - h_i  \cdot \langle \nicevec{\sigma}_i , \nicevec{e_2} \rangle=\mp 2 \cos \Big({\theta_i -\frac{\pi}{2}}\Big) = \mp 2 \sin{\theta_i}
\end{equation}

{Added to the attracting N/S effects on odd sites due to (\ref{HamPerio}), this periodic field   yields ground states and low-temperature states that are thus either a mostly East configuration $\theta^{ME}$, alternating periodically between NE, E, SE (selected {\em e.g.} when subject to an external all E field) or mostly West one $\theta^{MW}$, alternating NW, W, SW (if subject to an external all W field).}

\subsubsection{Zero-Temperature case}
The case $T=0$ we consider corresponds to infinitely strong bonds  but finite field strength.
\begin{lemma}\label{0TDisc}
The $2d-XY$ model with alternating field (\ref{Period}) on the decorated lattice $S=(2\Z^2)^c$ displays a spin-flop transition at zero temperature in the East-West direction ({perpendicular} to the field) : there {exist} two distinct ground states $\theta^{ME}$ and $\theta^{MW}$ with in particular opposite horizontal projections : $$\nicevec{\sigma}^{(1)}(\theta^{ME})=- \nicevec{\sigma}^{(1)}(\theta^{MW})\neq 0.$$
\end{lemma}

To see this, select, for some small $\epsilon >0$, the neighbourhoods $\mathcal{N}^{ME}=\mathcal{N}^{ME}_{\Lambda',\Delta',\epsilon}(\nicevec{\sigma}'_{\rm dblyalt})$ and $\mathcal{N}^{MW}=\mathcal{N}^{+\pi}_{\Lambda,\Delta,\epsilon}(\nicevec{\sigma}'_{\rm dblyalt})$, and use the global specification as in \cite{DDAELN} to show that, for $\nicevec{\omega}'^{ME} \in \mathcal{N}^{ME}$ and $\nicevec{\omega}'^{MW} \in \mathcal{N}^{MW}$, the expectations of the first coordinates are such that
\be \label{essdicvec}
 \Big| \nu \left[\nicevec{\sigma}'^{(1)}_{\{0,0\}} \mid \mathcal{F}'_{\{0,0\}^c}\right] (\nicevec{\omega}'^{ME})- \nu\left[\nicevec{\sigma}'^{(1)}_{\{0,0\}} \mid \mathcal{F}'_{\{0,0\}^c}\right] (\nicevec{\omega}'^{MW}) \Big| \;>\;  \delta
 \ee

Non-quasilocality  {\bf at zero temperature} follows directly.

Consider the first case, with a configuration $\sigma'_{{\rm dblyalt};\Delta' \setminus \Lambda'}{\nicevec{e}_{1}}_{\Delta'^{c}}$ doubly alternating configuration in the annulus $\Delta' \setminus \Lambda'$, but pointing {slightly East} outside $\Delta'$. The constrained Hamiltonian becomes a perturbation of (\ref{HamPerio}), with an external field becoming {mostly East directed but alternating NE or SE between two even spins on the annulus}, and zero outside:

\begin{eqnarray}\label{HamPerioE}
- \Phi_{i,j}^J(\nicevec{\sigma}) = J  \;\cdot  \langle \nicevec{\sigma}_i  , \nicevec{\sigma}_j \rangle  \mathbf{1}_{|i-j|=1} & \pm & 2  \cdot (\sin{\theta_i}-\sin{\theta_j}) \mathbf{1}_{i,j \in \Lambda, i=2k, j=2k'} \cdot \nicevec{e_2} \\ &{+} & 4  \cdot \mathbf{1}_{i \in \Delta, i=2k} {\cdot ({\pm} \nicevec{e_2} + \nicevec{e_1})}.\label{extrafield}
\end{eqnarray}

 Thus, due to the extra ``positive'' field (\ref{extrafield}) on the annulus $\Delta$, this Hamiltonian has {now only one ground state, the alternating mostly  East configuration} $\theta^{ME}$, so that here
 
\begin{equation}\label{SpinFlopEastZeroT}
\nu \left[\nicevec{\sigma}'^{(1)}_{\{0,0\}} \mid \mathcal{F}'_{\{0,0\}^c}\right] (\nicevec{\omega}'^{ME}) = \Gamma \big[ \nicevec{\sigma}^{(1)}_{\{0,0\}} \mid \omega^{ME} \big] = \; > \; 0.
\end{equation}

For  the other ground states (mostly West), selection is performed by conditioning on the neighbourhood 
$\mathcal{N}^{MW}$, leading   to a ``negative'' horizontal contribution in (\ref{extrafield}), so that
\begin{equation}\label{SpinFlopWestZeroT}
\nu \left[\nicevec{\sigma}'^{(1)}_{\{0,0\}} \mid \mathcal{F}'_{\{0,0\}^c}\right] (\nicevec{\omega}'^{MW}) = \Gamma \big[ \nicevec{\sigma}^{(1)}_{\{0,0\}} \mid \omega^{MW} \big] = \; < \; 0.
\end{equation}

Eventually, this discrete symmetry breaking leads to a spin-flop transition for the constrained model at zero temperature and to the essential discontinuity (\ref{essdicvec}).

\subsubsection{Non-Gibbs at low Temperature}

\begin{lemma}\label{LowTDisc}
The $2d-XY$ model with alternating field (\ref{Period}) on the decorated lattice $S=(2\Z^2)^c$ displays a spin-flop transition at low temperatures in the East-West direction (opposed to the field) : there exists two distinct Gibbs measures $\mu^{ME}$ and $\mu^{MW}$ such that in particular the expectations of the first coordinate satisfy $$\mu^{ME} [\nicevec{\sigma}^{(1)}]=-\mu^{MW} [\nicevec{\sigma}^{(1)}]\neq 0.$$
\end{lemma}

{\bf Proof:}
To get  this Lemma \ref{LowTDisc}, we need to investigate and prove the stability of the Ground states obtained  with the Hamiltonian (\ref{HamPerio}) with periodic field (\ref{Period}). {To do so, we have (at least) two options, depending on the context (mainly {depending on whether the interaction is n.n.~{\em vs.}~long-range, or whether the model is Reflection-positive or not})}.

In this $2d$-n.n. case, we follow the lines of continuous percolation of low energy ocean (item 1. below) while for more general models we indicate how to use more general contour arguments (item 2. afterwards)

\begin{enumerate}
\item{{\em Percolation of spin patterns}}, in the vein of Georgii~\cite{Geo81}, also Chapter 18 in \cite{Geo88}.

On the contrary to the {$2d$} classical $XY$/plane rotator model where the {uncountable} number of ground states prevents the existence of infinite percolation of spins in   favorized directions (see {\em e.g.}~Proposition 1.8 in \cite{Geo81}), the existence of two symmetric ground states in our model with periodic field allows it, leading afterwards to extremal symmetric and different measures supported by  clusters in opposite E/W directions.

From Theorem 3.7 in \cite{Geo81} (symmetry breaking in case of typical 
``small deviations clusters'' at low T), combined with Theorem 2.18 (Existence of unique infinite low-energy cluster in $2d$, n.n.), we adapt these arguments  in the next subsection  to get the following symmetry-breaking result for our $XY$ model with periodic external field (\ref{Period}). As we describe there, there is no major difference in the proof, we only have to incorporate the inhomogeneity of the self-interaction part due to the periodic field (depending on the site $i$ in relation to the period) and to consider larger $8*8$ blocks to restore some translation-invariance and {reflection-positive} properties.

For a given edge $v=\{i, j \}$ of the decorated lattice $S$, write $\phi_{ij} : E \times E \longrightarrow \R$ as
\begin{equation}\label{phiij}
\phi_{ij} (\nicevec{x},\nicevec{y}) = - J \langle \nicevec{x} ,  \nicevec{y} \rangle - h_i \langle \nicevec{x} ,  \nicevec{e_2} \rangle - h_j \langle \nicevec{y} , \nicevec{e_2} \rangle
\end{equation}
 such that formally
$$
H(\nicevec{\sigma})=\sum_{\{i,j\}} \phi_{ij}(\nicevec{\sigma}_i,\nicevec{\sigma}_j)\; ,
$$

so that  
\begin{equation}\label{eq.infabs}
m \coloneqq  {\inf_{i,j}} \inf_{(\nicevec{x},\nicevec{y})} \phi_{i,j}(\nicevec{x}, \nicevec{y})= -1\; 
\end{equation}
the latter sup being attained by the ground states $\theta^{ME}$ and $\theta^{MW}$ {\em via} the will of alignment of $\nicevec{\sigma}_i$ and $\nicevec{\sigma}_j$ with the alternating fields.

To get our low-temperature results with the same type of ingredients as the Peierls argument for the Ising model (which are, as recalled in \cite{Geo81}, percolation, symmetry, ground state stability), one needs first to adapt the graph of percolation clusters in the same way as we have adapted the topological neighbourhoods in the previous section: For any configuration $\nicevec{\sigma}$, for a given (small) $\epsilon>0$, consider the subgraph $G_\epsilon(\nicevec{\sigma})$ of low interaction bonds
$$
G_\epsilon(\nicevec{\sigma}) \coloneqq \big\{ \{i,j \} \in L,\phi_{ij}(\nicevec{\sigma}_i, \nicevec{\sigma}_j) \leq m
+ \epsilon \big\}\; ,
$$
where $L$ is the set of (n.n.) edges of the decorated lattice $S$.

To get the searched-for symmetry breaking, we need to investigate the existence of an infinite connected component in this graph $G_\epsilon$, the event written $\{\exists C_\infty^\epsilon \}$ (while the existence of a unique such cluster is written 
$ \{ \exists 1 C_\infty^\epsilon \}$). Note that such a percolation holds when the spins have almost all the same orientation and that the latter is close enough to the ground states $\theta^{ME}$ or $\theta^{MW}$ (as this requires a deviation from the ground states smaller than $\epsilon$). By adapting the techniques of~\cite{Geo81} with an extra alternating field, we first get that such an existence and uniqueness is typical at low temperatures. As there are two such types of clusters, each of the  measure obtained by pointed either mostly East or either mostly West will have its own typical cluster. For this model, one gets that there {is} discrete symmetry breaking using the following two facts:
\begin{description}
\item{\bf Two symmetric ground states:} Our Hamiltonian (\ref{HamPerio}) admits two ground states $\theta^{ME}$ and $\theta^{MW}$, {related by a flip of all the spins along} the vertical axis.
\item{\bf Reflection-Positivity:}   see {\em e.g.} Lemma 10.8 or Example 10.9 in \cite{FV16}, also \cite{FILS78,FL78}.
\end{description}

{To fall exactly within this framework of Georgii \cite{Geo81}, we divide the lattice $\mathbb{Z}^2$ into blocks $B_{i,j}$ of size 8 by 8 such that the block with coordinate $(i,j)$ has its center at the site  $(5+8i, 5+8j)$.
The energy term of a block contains all bonds and sites inside the block, plus half the contribution of the boundary bonds. Note that the Hamiltonian is invariant under translations of distances which are multiples of $8$. Notice also that we can incorporate the periodic fields from the constraints in the a priori measure of the block spins. 
Moreover the system  is reflection invariant and Reflection Positive for reflecting about the planes (either horizontal or vertical) at distance $4$ from each other in which either $x$-coordinate or $y$-coordinate is of the form $4n +3$, with $n\in \mathbb{Z}$.
}

 More precisely, we build a $8\times 8$ block system as follows: Consider new block-spin variables $B=\cup_{x \in 8 \mathbb{Z}^{2}}B_{x}$. For a spin configuration $\nicevec{\sigma} \in \Omega$,  denote a generic image configuration by $X=X(\nicevec{\sigma}) \in E^B$. Then, the interaction among blocks $B_{x}$ and $B_{y}$ is
\begin{equation}
\label{PsixyXX}
\begin{split}
\Psi_{xy}(X)&= -\sum_{\substack{i \sim j \\ i,j \in B_{x}}} \phi_{ij}( \nicevec{\sigma}_{i},\nicevec{\sigma}_{j})- \sum_{\substack{k \sim l \\ k,l \in B_{y}}}\phi_{ij}(\nicevec{\sigma}_{k} , \nicevec{\sigma}_{l} )- \frac{1}{2} \sum_{\substack{i \in \partial B_{x}\\ j \in \partial B_{y}\\ i \sim k}}\phi_{ij}(\nicevec{\sigma}_{i} , \nicevec{\sigma}_{k} )
 \end{split}
\end{equation}
It is a $C$-potential in the sense of Georgii. For $n \in \mathbb{Z}$, consider translations by $8$ in the two coordinate directions, i.e.~
$$
\left[\tau^{(n)}_{k}(\nicevec{\sigma})\right]_{v} \overset{\text{def}}{=} \nicevec{\sigma}_{v + 8n \nicevec{e}_{k}}, \qquad v\in \mathbb{Z}^{2}, n \in \mathbb{Z},
$$ 
and the family of reflections along any axis passing through type 1 or 2 spins, namely
$$
\left[r^{(n)}_{k}(\nicevec{\sigma})\right]_{v} \overset{\text{def}}{=}  \nicevec{\sigma}_{v-(4n+1) \nicevec{e}_{k}}\qquad v\in \mathbb{Z}^{2}, n \in \mathbb{Z},
$$
which are obviously involutive, i.e.~$r^{(n)}_{k}\circ r^{(n)}_{k}= \text{Id}$, for $k=1,2$. Clearly, potential (\ref{PsixyXX}) is translation invariant w.r.t.~$\tau^{(n)}_{k}$, $k=1,2$, for all $n \in \mathbb{Z}$. Moreover, being a sum of scalar products, it is also Reflection-Positive w.r.t.~$r^{(n)}_{k}$ ($\forall n$): For any local function $f$,
$$
\mu[f \cdot f^* ] >0
$$
where $f^{*}=f \circ r^{(n)}_k$, $k=1,2$.

Following  the context of \cite{Geo81}, Section 3, Theorem 3.7, one gets the validity of a chessboard estimate to derive afterwards an ad-hoc Peierls estimate. This allows us to establish the existence of two distinct symmetric and mutually singular  Gibbs measures, supported by the sets of configurations with a unique ``low-energy cluster'', which can be seen as low-temperature perturbation of our two ground states, roughly pointing in one of the favorite directions, mostly East or mostly West.

Let us describe how our model fits in the framework of this method, that uses the usual ingredients to implement a Peierls-type argument: Percolation (clusters $\mathcal{C}^\epsilon_\infty$ of "small deviation" vector), symmetry and ground states stability. 

\begin{theorem} {\em (From Thm 2.18 in \cite{Geo81})}\label{Unique} For $\epsilon >0$ small enough, at low enough temperature, there is at least one {(periodic)} $\mu \in \mathcal{G}(\gamma^J)$ such that
$$
\mu( \exists C_\infty^\epsilon )=\mu(\exists 1 C_\infty^\epsilon)=1.
$$
\end{theorem}

The measure used in \cite{Geo81} is the Gibbs measure with periodic boundary conditions but the existence of any such a measure is enough: from this result, one can afterwards associate a Gibbs measure by a perturbation of the ground states $\theta^{ME}$ and $\theta^{MW}$ -- by conditioning on the different typical configurations with the two types of small deviation clusters, see below -- and get two different ones at low temperatures, pointing globally mostly East or mostly West. Recall that for us in this section, the state space is still $E=\mathcal{S}^1$ while the  lattice is the decorated lattice $S=(2 \mathbb{Z}^2)^c$, where $L$ is the set of its n.n.~edges. 

\begin{theorem}{\em (From Thm 3.7 in  \cite{Geo81})}\label{SB} {At low enough temperature}, the set of Gibbs measures for the Hamiltonian (\ref{HamPerio}) {contains} (at least) two elements supported {on} two disjoint sets $A_{ME}$, $A_{MW}$ supporting low energy  clusters, for $\epsilon >0$, so that  
$$
\big\{\nicevec{\sigma} \in E^L : \sup_{\langle ij \rangle \subset L}  \Phi_{ij}(\nicevec{\sigma}) \leq m + \epsilon \big\}  = A_{ME} \cup A_{MW}
$$

This yields Gibbs measures $\mu^{ME}$ and $\mu^{MW}$ from Theorem \ref{MW}  with the connected components $C_{ME}$ and $C_{MW}$ such that 
$$
\mu^{ME}(C_{ME})=\mu^{MW}(C_{MW})=1.
$$
\end{theorem}

The {low-energy} clusters $C_{MW}$ and $C_{MW}$ are also called small deviation clusters. As soon as this infinite percolation clusters {occur with non-zero probability} w.r.t.~to the periodic {measure} $\mu$, one gets
\begin{equation}\label{MuEMuW}
\mu^{ME}(\cdot) = \mu ( \cdot \mid C_{ME}) \; {\rm and} \; \mu^{MW} (\cdot) = \mu ( \cdot \mid C_{MW}).
\end{equation}

This {will} lead  to the opposite inequalities (\ref{SpinFlopEast}) and (\ref{SpinFlopWest}) below.

{\bf Consequence: Spin-flop transition and essential discontinuity}

To make use of these two symmetric extremal measures, we remark that they indeed point  typically in opposite directions  (this can be seen from (\ref{MuEMuW}) and the structure of the clusters) to get on the same sub-neighbourhoods $\mathcal{N}^{ME}$ and $\mathcal{N}^{MW}$ as in the {zero-temperature cases}, the two opposite mean orientations, towards East in the Eastern boundary orientation,
\begin{equation} \label{SpinFlopEast}
\forall \nicevec{\omega}'^{ME} \in \mathcal{N}^{ME},\; \nu \left[\nicevec{\sigma}'^{(1)}_{\{0,0\}} \mid \mathcal{F}'_{\{0,0\}^c}\right] (\nicevec{\omega}'^{ME}) = \mu^{ME} \big[\nicevec{\sigma}'^{(1)}_{\{0,0\}} \big] \;> \;  0
\end{equation}
while the other infinite cluster (towards West) is selected for conditionings on the second neighbourhood 
$\mathcal{N}^{MW}$, to get

\begin{equation} \label{SpinFlopWest}
\forall \nicevec{\omega}'^{MW} \in \mathcal{N}^{MW},\; \nu \left[\nicevec{\sigma}'^{(1)}_{\{0,0\}} \mid \mathcal{F}'_{\{0,0\}^c}\right] (\nicevec{\omega}'^{MW}) = \mu^{MW} \big[ \nicevec{\sigma}'^{(1)}_{\{0,0\}}\big] \; < \; 0.
\end{equation}

Eventually, this leads to a spin-flop transition at low  temperature {\em via} this discrete symmetry breaking and to the essential discontinuity (\ref{essdicvec}).

 This proves Theorem \ref{DecXY} for the temperatures below the percolation temperature for small percolation cluster (depending on the deviation bound $\epsilon$).

{\bf Remark:} Note that this method to prove the occurrence of a spin-flop transition makes use of Reflection-Positivity (for the existence of a unique infinite cluster for the Gibbs measure with periodic-b.c., in Theorem 4), and could thus be adapted to other models enjoying this RP property, such as long-range models considered in {the} next section, but could not be applied to other models not enjoying it, as {\em e.g.} next-to-nearest-neighbours models. Nevertheless, in addition to RP, this low-energy-ocean method of Georgii is derived for n.n.~models and would require an adaptation to long-range RP models, as shortly discussed in {the} next sections.

\item{{\em Generalized contour models}} from Malyshev {\em et al.}~\cite{MMPT}, following the lines of  \cite{EKO2011}.

Our Hamiltonian (\ref{HamPerio}) satisfies the hypotheses of a {general} Theorem by Malyshev {\em et al.} (\cite{MMPT}, Theorem 6). Roughly speaking, this theorem of  \cite{MMPT} states that if the Hamiltonian a) admits $N$ different global minima and b) is sufficiently regular and satisfies a certain local convexity property around those $N$ minima, then the existence of those $N$ minima (i.e.~the $N$ distinct {zero-temperature} Gibbs measures) \emph{persists} at finite (sufficiently small) temperature. The argument is along the lines of~\cite{EKO2011} and goes as follows. 

First, one needs to (slightly) modify the potential and build a nonnegative function $\Psi$ of the joint variables in the product space $[0,\pi] \times [0,\pi] \times \mathbb{R}^{+}$, the latter space representing the temperature, the unique parameter in our case. We call such a function $\Psi: [0,\pi] \times [0,\pi]  \times  \mathbb{R}^{+} \to \mathbb{R}$. It is defined, at zero temperature, by
\be
\label{eq.modpot}
\Psi(\theta,\omega;0) \coloneqq \Phi(\theta)+\Phi(\omega) + c \; ,
\ee
where $c=-2m$ and $m=\inf \Phi$,  see (\ref{eq.infabs}). Clearly $\Psi$ has $N=2$ distinct minima, located at points $(\theta,\omega)=(0,0)$ and $(\theta,\omega)=(\pi,\pi)$ (corresponding, respectively, to the zero temperature Gibbs measures $\mu^{+}$ and $\mu^{-}$).  Also, by definition $\Hessian\Psi$ is block diagonal (i.e.~$\frac{\partial^{2}}{\partial \theta \partial \omega}\Psi(\theta,\omega;0)=0$) and thus, in order to have the Malyshev \emph{et al.} persistence result in our special case, third condition of~\cite{MMPT}, it is sufficient to show that $\det\left(\Hessian{\Phi}\right)\big |_{(\theta,\omega) = (0,0),(\pi,\pi)}>0$. Due to the particular form of ( \ref{eq.modpot}), this follows if 
$$
\frac{\partial^2\Phi(\theta,\omega;0)}{\partial \theta_{i}^{2}}\big |_{\theta_{i} = 0,\pi}=\frac{\partial^2\Phi(\theta,\omega;0)}{\partial \omega_{i}^{2}}\big |_{\omega_{i} = 0,\pi}
$$ 
is non-null. But actually, in our case the \ref{eq.modpot} is bounded away from zero, more precisely
\be
\label{eq.curv}
\frac{\partial^2\Phi(\theta,\omega;0)}{\partial \theta_{i}^{2}}\big |_{\theta_{i} = 0,\pi} = J \sum_{i \sim j} \cos{(\theta_{i}-\theta_{j})}\big |_{\theta_{i} = 0,\pi} \pm \sin{\theta_{i}}\big |_{\theta_{i} = 0,\pi}= \pm 4J,
\ee
and hence, by Theorem 6 in~\cite{MMPT}, if $\beta$ is large enough there exists at least two distinct Gibbs measures for $\Psi$, and hence for our potential $\Phi$. 

The Malyshev {\em et al.} contour method can be extended to long-range models by softening contours. However, the percolation techniques used in this $2d-n.n.$ framework, which give chessboard estimates and hence the Peierls argument, can also be extended in this case along the same lines. We shall pursue the latter path, and leave the former (i.e.~the softening of these contours) to a future work.

\end{enumerate}

\section{Decimation of the borderline $1/r^{2d}$- long-range  rotators} 

We briefly describe two extensions for long-range models:

{\bf 1. $1/r^2$-Long-range rotator model in $1d$}\label{LR1d}

Consider long-range pair potentials of the form of (\ref{XY}),
\be \label{XY2}
\Phi_A(\sigma)=\left\{
\begin{array}{lll}
\; - J(i,j) \;\cdot  \langle \nicevec{\sigma}_i , \nicevec{\sigma}_j \rangle \; \; & &\textrm{if} \; A=\{i,j\} \;\\
\; 0 \; & &\textrm{otherwise}
\end{array} \right.
\ee
where $\langle \cdot , \cdot \rangle $ is still {\em some} inner product in $\R$, as the configurations still takes values on the sphere $\mathbb{S}^1$, with  one-dimensional  long-range couplings
$$
J(i,j)=\frac{J}{|i-j|^2}, \; J \geq 0.
$$

An analogous version of MW Theorem, with Ferromagnetic uniqueness as in Theorem \ref{MW}, has been proven by Simon in \cite{Sim81}, while  the analogous discrete Ising version has been well investigated since the late sixties (see    Kac  {\em et al.} \cite{KT69}, Dyson \cite{Dys71},  Fr\"ohlich {\em et al.} \cite{FrSp81, FrSp82} or   Imbrie {\em et al.} \cite{I82,IN88}) and is  known to present a symmetry-breaking phenomenon, leaving the door open to a spin-flop phenomenon if one follows the lines of the preceding section.

In this long-range context, it appears, as often, {that} the one-dimensional picture {resembles} more the $2d-n.n.$ case: considering exactly the same settings as in Section 3, replacing $d=2$ by $d=1$ in the configuration space and definition of Hamiltonian (\ref{HamPerio}) with alternating fields (\ref{HamPerioE}) or (\ref{Period}). Indeed, while the simply alternating configuration is good due to exact {cancellations}, the doubly alternating $...NNSSNNSS...$ one on the decimated lattice yields a similar diluted alternating field. as we briefly indicate in the introduction, only the value of the field is depressed by a scaling constant $2G$ where $G$ is the Catalan number.

{We note that the constrained invisible-spin model has an interaction which correlates the spin coordinates in the EW direction, and which decays like $1/r^2$. Our spins are ``soft'', not two-valued, but to extend combinatorial, contour-type arguments to a discrete symmetry-breaking of a soft-spin model is what the arguments of Malyshev {\em et al.}~\cite{MMPT} manage. For another example illustrating the robustness of the transition for interactions decaying as $1/r^2$, see also Fontes~\cite{Fontes93}.}

{\bf 2. $1/r^4$-Long-range rotator model in $2d$}\label{LR2d}

In  contrast to the preceding item, in $2d$ for  the {critical} long-range {intermediate} phase in $1/r^4$, there are not so many cancellations of sites and the external field (\ref{Period}), due to the merging of other types of sites (7). Some are still zero but inside squares the sites experience a field, smaller than the aligned alternation of sites between odd/even sites, with the same period. Nevertheless, {as the decay is faster than {$r^{-3}$}, ordinary contour arguments can be used to prove the existence of a transition in the conditioned invisible-spin system \cite{Ger76}.}

\section{Comments and conclusions}

{Decimation transformations are  in some sense the simplest version of Renormalisation Group maps. 
It is hoped that their behaviour illustrates properties which may apply more widely.\\
However, there arise a number of problems if one wants to implement the physics {ideas} by invoking results about decimation transformations. The first requirement of existence of the transformation already turns out to be non-trivial.  Indeed, non-Gibbsianness of a transformed measure means non-existence of a renormalised interaction, inside a relatively large interaction space. \\
 One noteworthy question appears to be the relation between the existence of phase transitions in an original model versus the non-Gibbsianness of the transformed measures. 
 When the system is deep in the uniqueness region,  decimation maps look well-behaved, as decimated measures tend to be Gibbsian, \cite{GP79, Isr81}. \\
In~\cite{EFS93,ELN17, DDAELN} it was shown that at very low temperatures decimated measures are non-Gibbsian, in the phase transition region, or close {to} it. The work of \cite{MO} indicates that iterating the transformation moves the region of non-Gibbsianness towards  the whole coexistence region. \\
Around the transition temperature, Ising models possess a region, including part of the coexistence region, where  decimated measures are Gibbs measures \cite{HalKen,Ken}.  However, if one tries the same  for Potts models, the results are of opposite nature,  and the transition point lies inside a region where the decimated measures are non-Gibbsian \cite{EFK}.\\
{In this work we concentrated on 2-component vector spins (classical $XY$ spins).} We found a new source of non-Gibbsian examples, in models which in its original form do not display long-range order at any temperature or field, but which nonetheless become non-Gibbsian after decimation. The underlying mechanism rests on the fact that by conditioning one reduces the symmetry of the model from continuous to discrete, which allows to make use of the spin-flop transition for conditioned measures. \\
As a last remark, we notice that even if a transformed interaction exists and the decimation transformation is well-defined, it still may not behave as {physical intuition} would predict in different (spectral) aspects \cite{Yin}.

 }

{\bf Acknowledgments:} 
We thank the referee for some helpful remarks. Our research {has} been partially carried out  within the CNRS IRP B\'ezout-Eurandom ``Random Graph, Statistical Mechanics and Networks'', supported by Laboratory LAMA (CNRS UMR8050), B\'ezout {Federation} (CNRS Unit FR3522), Labex B\'ezout (ANR-10-LABX-58) and Eurandom (TU/e Eindhoven).


\addcontentsline{toc}{section}{\bf References}
\begin{thebibliography}{23}




\bibitem{ACCN} M. Aizenman, J.T. Chayes, L. Chayes, C.M. Newman. Discontinuity of the magnetization in one-dimensional $\frac{1}{{|x-y|}^{2}}$ Ising and Potts models {\em J. Stat. Phys.} {\bf 50}: 1--40, 1988.  

\bibitem{BGMMT21} S. Barbieri,  R. G\'omez, B. Marcus, T. Meyerovitch, S. Taati.  Gibbsian Representations  of Continuous Specifications: The Theorems of Kozlov and Sullivan Revisited,  {\em Comm. Math. Phys. } {\bf 382}:  1111-1164, 2021.



\bibitem{BFL77} J. Bricmont, J.-R. Fontaine, L. Landau. On the Uniqueness of Equilibrium State for Plane Rotators. {\em Comm. Math. Phys.} {\bf 56}:281--290, 1977.


 
 
\bibitem{CFMP} M. Cassandro, P.A. Ferrari, I. Merola, E. Presutti. Geometry of contours and Peierls estimates in $d=1$ Ising models with long range interactions. {\em J. Math. Phys. } {\bf 46}: 053305, 2005. 
 
\bibitem{CR2016} F. Collet, W.M. Ruszel. Synchronization and Spin-Flop Transitions for a Mean-Field $XY$-Model in Random Field. {\em J. Stat. Phys.} {\bf 164}:645--666, 2016.



\bibitem{Craw2011} N. Crawford. On Random-Field Induced Ordering in the Classical $XY$ Model. {\em J. Stat. Phys.} {\bf 14}, no 1:11--42, 2011.

\bibitem{Craw2014} N. Crawford. Random Field induced order in low dimension I. {\em Comm. Math. Phys.} {\bf 328}, no 1:203--249, 2014.

\bibitem{CrawRuszel} N. Crawford, W. Ruszel. Random Field induced order in Two Dimensions. Preprint arXiv:2111.00241v1, 2021.

\bibitem{DDAELN} M. D'Achille, A. van Enter, A. Le Ny. Decimations for two-dimensional Ising and rotator models. {\em J. Math. Phys.}~{\bf 63}: 033506, 2022.

\bibitem{Dob68} R.L.~Dobrushin. The Description of a Random Field by Means of Conditional Probabilities and Conditions of its Regularity.
{\em Th. Prob. Appl.} {\bf 13}:197--224, 1968.
	



	

\bibitem{Dys71} F. Dyson. An Ising ferromagnet with discontinuous long-range order. {\em Comm. Math. Phys.} {\bf 21}:269--283, 1971.



 
\bibitem{EFS93} A.C.D. van Enter, R.~Fern{\'a}ndez, A.D.  Sokal.  Regularity Properties and Pathologies of Position-Space R.G. Transformations: Scope and Limitations of Gibbsian Theory.
 {\em J. Stat. Phys.} {\bf 72}: 879-1167, 1993.

\bibitem{EFK} A.C.D. van Enter, R. Fern{\'a}ndez, R. Koteck\'y. Pathological behavior of renormalization-group maps at high fields and above the transition temperature. {\em J. Stat. Phys.} {\bf 79},969--992, 1995.  


\bibitem{EKO2011} A.C.D. van Enter, C. K\"ulske, A.A.  Opoku. Discrete Approximations to Vector Spin Models. {\em J. Phys. A} {\bf 44}, no 47, 475002, 2011.

\bibitem{EKOR2010} A.C.D. van Enter, C. K\"ulske, A.A.  Opoku, W. M. Ruszel.
Gibbs-non-Gibbs Properties for $N$-vector Lattice and Mean-Field Models.  {\em Braz. J. Probab. Stat.} {\bf 24},  no. 2, 226–255, 2010.

\bibitem{ELN17} A.C.D. van Enter, A. Le Ny. Decimation of the Dyson-Ising Ferromagnet. {\em Stoch. Proc. Appl.} {\bf 127}:  3776--3791, 2017.

\bibitem{ER2008} A.C.D.  van Enter, W.M. Ruszel. Loss and Recovery of Gibbsianness for $XY$-Models in External Fields. {\em J. Math. Phys.}{\bf 49}:125208,  2008.

\bibitem{ER2009} A.C.D.  van Enter, W.M. Ruszel. Gibbsianness {\em vs.} non-Gibbsianness of Time-Evolved Planar Rotor Models. {\em Stoch. Proc. Appl.} {\bf 119}:1866--1888, 2009.

 

 
 \bibitem{FP97} R. Fern\'andez, C.-E. Pfister. Global specifications and Non-Quasilocality of Projections of Gibbs Measures. {\em Ann. Prob.} {\bf 25}, no 3:1284-315, 1997.

 
\bibitem{Foll75} H. F\"ollmer. Phase Transition and Martin Boundary. In {\em S\'eminaires de Probabilit\'es IX, Universit\'e de Strasbourg}. Lecture Notes in Mathematics {\bf 465}:305--317, Springer, 1975.


\bibitem{Fontes93} {L.~Fontes. An Ordered Phase with Slow Decay of Correlations in Continuum $1/r^2$ Ising Models. {\em Ann.~Prob.} {\bf  21}, no. 3, pp. 1394–412, 1993.}
 
\bibitem{FV16} S.~Friedli, Y.~Velenik. Statistical Mechanics of Lattice Systems: a Concrete Mathematical Introduction. 
Cambridge University Press, 2017.

\bibitem{FILS78} J. Fr\"ohlich, R.B. Israel, E.H. Lieb, B. Simon. Phase Transitions and Reflection Positivity. I. General theory and Long-Range Lattice Models. {\em Comm. Math. Phys.} {\bf 62}:1-34, 1978.

\bibitem{FL78} J. Fr\"ohlich, E.H. Lieb. Phase Transitions in Anisotropic Lattice Spin Systems. 
 {\em Comm. Math. Phys.} {\bf 60}:233--267, 1978.
 

\bibitem{FrSp81} J. Fr\"ohlich, T. Spencer.  The Kosterlitz-Thouless in Two-Dimensional Abelian Spin Systems and the Coulomb Gas. 
{\em Comm. Math. Phys.} {\bf 81}:527--602, 1981.

\bibitem{FrSp82} J. Fr\"ohlich, T. Spencer.  The Phase Transition in the One-Dimensional Ising Model with $1/r^2$ Interaction Energy. 
 {\em Comm. Math. Phys.} {\bf 84}:87--101, 1982.
 


\bibitem{FrPf81} J. Fr\"ohlich, C.-E. Pfister. On the Absence of Spontaneous Symmetry Breaking and of Crystalline Ordering in $2d$ Systems.
{\em Comm. Math. Phys.} {\bf 81}:277-298, 1981.

\bibitem{FrPf83} J. Fr\"ohlich, C.-E. Pfister. Spin Waves, Vortices, and the  Structure of Equilibrium States in the Classical $XY$-Model.
{\em Comm. Math. Phys.} {\bf 89}:303--327, 1983.

\bibitem{Ger76} V.~Gertzik. Conditions of the nonuniqueness of Gibbs state for lattice models with Finite Potential.
{\em Izvestija Akademii Nauk. SSSR Ser. Matem.}  {\bf 40}:448-462, 1976. 


\bibitem{Geo81} H.-O. Georgii. Percolation for Low Energy Clusters and Discrete Symmetry Breaking in Classical Spin Systems. {\em Comm. Math. Phys.} {\bf 81}:455--473, 1981. 

\bibitem{Geo88} H.-O. Georgii. {\em Gibbs Measures and Phase Transitions}.
De Gruyter Studies in Mathematics {\bf 9} Berlin--NY, 1988.
See also $2^d$ edition, 2011.

\bibitem{GP79} R.B. Griffiths, P.A. Pearce. Mathematical properties of position-space renormalization-group transformations. {\em J. Stat. Phys.} {\bf 20}: 499--545, 1979.

\bibitem{HalKen} K. Haller, T. Kennedy.  Absence of renormalization group pathologies near the critical temperature. Two examples. {\em J. Stat. Phys.} {\bf 85}: 607--637, 1996. 

\bibitem{I82} J. Imbrie. Decay of correlations in one-dimensional Ising model with $J_{ij}=\mid i-j \mid^{-2}$.
{\em Comm. Math. Phys.} {\bf 85}:491--515, 1982.
 
 \bibitem{IN88} J. Imbrie, C.  Newman. An intermediate phase with slow decay of correlations in one-dimensional $1/\mid x-y \mid^2$ percolation, Ising and Potts Models.
{\em Comm. Math. Phys.} {\bf 118}:303--336, 1988.

	
\bibitem{Isr81} R.B. Israel. Banach Algebras and Kadanoff Transformations. In {\em Random Fields (Esztergom, 1979)} J. Fritz, J.L. Lebowitz and D. Sz\'asz eds, vol II, pp 593-608, 1981. 

\bibitem{KT69} M.~Kac,  C. Thompson. Critical behavior of several lattice models with long-range interaction. 
{\em J. Math. Phys.} {\bf 10}:1373--1386, 1969.

\bibitem{Ken} T. Kennedy. Absence of renormalization group pathologies in some critical Dyson-Ising ferromagnets. arXiv:2006.11429, 2020.

  \bibitem{Ko} O.K.  Kozlov. Gibbs Description of a System of Random Variables. {\em Problems Inform. Transmission.} {\bf 10}:258--265, 1974.
  





\bibitem{LR69}O.E.~Lanford, D.~Ruelle. Observables at Infinity and States with Short Range Correlations in Statistical Mechanics.  
{\em Comm. Math. Phys.} {\bf 13}:194--215, 1969.



\bibitem{LN08} A. Le Ny. {\em Introduction to Generalized Gibbs measures}. Ensaios Matem\'aticos {\bf 15}, 2008.


\bibitem{MMPT} V.A. Malyshev, R.A. Minlos, E.N. Petrova, Y.A. Terletskii. Generalized contour models. {\em The Journal of Mathematical Sciences} {\bf 23} 2501–2533, 1983.

\bibitem{MO} F. Martinelli, E. Olivieri. Some remarks on pathologies of renormalization-group transformations for the Ising model. {\em J. Stat. Phys. } {\bf 72}: 1169--1177. 

\bibitem{Mermin67} N.D. Mermin. Absence of Ordering in Certain Classical Systems. {\em J. Math. Phys.} {\bf 8}:1061--1064, 1967.

 \bibitem{MW1966} N.D. Mermin, H. Wagner. Absence of Ferromagnetism or Antiferromagnetism in One- or Two--dimensional Isotropic Heisenberg Models. {\em Phys. Rev. Lett.} {\bf 17}:1133--1136, 1966.

\bibitem{MMSP78} A. Messager, S. Miracle-Sole, C.-E. Pfister. Correlation Inequalities and Uniqueness of the Equilibrium State for the Plane Rotator Ferromagnetic Model. {\em Comm. Math. Phys.} {\bf 58}:19--29, 1978.


\bibitem{Monroe75} J.L. Monroe. Correlation Inequalities for Two-Dimensional Vector Spin Systems. {\em J. Math. Phys.} {\bf 16}:1809--1812, 1975. 



\bibitem{Pres80} C.~Preston. Construction of Specifications. In {\em Quantum Fields - Algebras, Processes (Bielefeld symposium 1978)} pp 269-282, ed. L. Streit, Wien-NY : Springer 1980.





\bibitem{Sim81} B. Simon.  Absence of continuous symmetry breaking in a one-dimensional 
$n^{-{{2}}}$ model. {\em J. Stat. Phys. } {\bf 26},  307-311, 1981.


 
 
\bibitem{Su} W.G. Sullivan. Potentials for Almost Markovian Random Fields.  {\em Comm. Math. Phys.} {\bf 33}:61-74, 1973.
 
 \bibitem{Yin} M.~Yin. Spectral properties of the renormalization group  at infinite temperature.  {\em Comm.Math.Phys.} {\bf 304}, 175--186,  2011.  



\end{thebibliography}
\end{document}